\documentclass[twocolumn]{aastex631}

\usepackage{amsmath}
\begin{document}

\title{An Extreme Scattering Event Toward PSR J2313+4253}

\author[0000-0001-6620-5752]{Zachary C. Zelensky}
\affiliation{Department of Physics and Astronomy, Texas Tech University, 
Lubbock,TX, 79410-1051, USA}

\author[0000-0002-2451-7288]{Jacob E. Turner}
\affiliation{Green Bank Observatory, 
 P.O. Box 2, Green Bank, WV, 24944, USA}

 \author[0000-0001-8233-5442]{Juan G. Lebron Medina}
\affiliation{Department of Physics, University of Puerto Rico at Mayagüez, 6V66+C8C, Mayagüez, 00680, Puerto Rico}

\author{Daniel E. Reichart}
\affiliation{Skynet Robotic Telescope Network, University of North Carolina, Chapel Hill, NC 27599, USA}

\author{Joshua B. Haislip}
\affiliation{Skynet Robotic Telescope Network, University of North Carolina, Chapel Hill, NC 27599, USA}

\author{Vladimir V. Kouprianov}
\affiliation{Skynet Robotic Telescope Network, University of North Carolina, Chapel Hill, NC 27599, USA}

\author{Steve White}
\affiliation{Green Bank Observatory, 
 P.O. Box 2, Green Bank, WV 24944, USA}

\author{Frank Ghigo}
\affiliation{Green Bank Observatory, 
 P.O. Box 2, Green Bank, WV 24944, USA}

\author{Sue Ann Heatherly}
\affiliation{Green Bank Observatory, 
 P.O. Box 2, Green Bank, WV 24944, USA}

\author[0000-0001-7697-7422]{Maura A. McLaughlin}
\affiliation{Department of Physics and Astronomy, West Virginia University, P.O. Box 6315, Morgantown, WV 26505,USA}
\affiliation{Center for Gravitational Waves and Cosmology, West Virginia University, Chestnut Ridge Research Building, Morgantown, WV 26505,USA}



\begin{abstract}

We present evidence of an extreme scattering event (ESE) toward PSR J2313+4253 using high-cadence observations taken with the Green Bank Observatory 20m telescope.  The high density of observations in time allow for detailed tracking of the event. We observe a pair of spikes along with the characteristic drop in scintillation bandwidth that is expected during an ESE. This pattern implies that the structures predominantly responsible for scattering occur at different distances than those from previous and subsequent epochs. A secondary spectrum processed during the event shows a detached feature similar to those found in double lensing events from previously observed ESEs. We measure this event as originating from a scattering region with a distance of 1.04(1) kpc, a transverse size of 15 AU, and a duration of approximately 220 days. These rare events provide opportunities to study the properties of small-scale structures in the ISM.

\end{abstract}



\section{Introduction} \label{sec:intro}
Pulsar scintillation is a useful tool for probing AU-parsec scale structures in the ionized interstellar medium (ISM). Scintillation originates from the multipath radio emission from a pulsar interacting with free electrons in the ISM at a scattering screen. The result of this interaction causes a phase shift as photons with differing path lengths are directed toward Earth. When this emission recombines at a telescope, an interference pattern is created that evolves in frequency and time. This interference pattern is seen in the pulsar's dynamic spectrum, which shows flux density as a function of frequency and time, and manifests as regions of constructive inference, known as scintles. Secondary spectra, which show a 2D Fourier transform of the dynamic spectra, can exhibit parabolic structures known as scintillation arcs \citep{Stinebring_2001}. Scintillation arcs arise from photons scattered by a thin screen at different incident angles interfering with photons taking a line-of-sight path. These arcs are useful probes for the structure and origin of the scattering screen and can be used to estimate the screen distance \citep{10.1093/mnras/stad3683}. If the arc is sharp, the scattering is dominated by a constrained region. In contrast, a fuzzy arc would indicate a more diffuse scattering region \citep{DanielR.Stinebring_2006}. Pulsar scintillation has been extensively reviewed in \cite{ricket1,ricket2}.

Extreme scattering events (ESEs), first observed as flux density variations of quasars, are theorized to occur from small-scale variations in the ISM \citep{1987Natur.326..675F}. These events are often identified by a trend observed in the flux density time series from a source. ESEs manifest as a frequency-dependent change in the observed flux, likely caused by a lens-like structure. The resulting flux density trend contains caustic spikes, created as rays pile up at the edges of the lens, and a dip in the center of the lens, as the rays are defocused. The magnitude of the flux decrease, as well as the distance between these spikes, can be useful in determining properties of the ESE source. This is especially true in cases where the event has been observed in high detail. ESEs have also been studied with VLBI imaging to detect angular broadening and  test models of the structural origin \citep{ESE_vlbi}. Real-time identification of ESEs has demonstrated the possibility of real-time multi-wavelength followup observations \citep{Bannister_2016}. New techniques have been developed to reconstruct plasma lens structures using dynamic spectra during scattering events. This allows for the screen properties and structure to be analyzed \citep{Tuntsov_2016}. 

Since their discovery, ESEs have also been observed in pulsars, with the first such event observed toward the millisecond pulsar B1937+21 \citep{1993Natur.366.320}. ESEs can be tracked in pulsar observations through several observables, including flux variations, changes in dispersion measure (DM), which is the integrated column density of free electrons along the line of sight, scintillation bandwidth and timescale, which are the characteristic width of the observed scintles in frequency and time, respectively, inferred scattering screen distance, or angular broadening measurements \citep{main_2022,Reardon2020}. Often a simultaneous change in two or more of these parameters is seen. For example, several ESEs were detected through correlated scintillation bandwidth and DM variations \citep{2015ApJ...808..113C}. Similarly, two ESEs were detected through flux density and scintillation bandwidth variations \citep{10.1093/mnras/stx3101}. These events can range in duration from several weeks to a few years, with a particularly long event observed as lasting three years \citep{2003ApJlongest}.

 The origin of these events is not fully understood; 
however, several models have been proposed. One such explanation models the ESE's source as a Gaussian plasma lens \citep{1998ApJ...496..253C}. Plasma lenses are diverging lenses; this creates a drop in intensity near the center of the lens and causes light rays to pile up at the edges, forming caustics. Simulations suggest that pulsar ESEs can be caused naturally by the turbulence in the ionized interstellar medium \citep{2007simulation}. Another explanation uses catastrophe theory to explain both pulsar scintillation and ESEs in a single, unified framework. In this model, pulsar scintillation occurs from \(A_2\) catastrophes, which are folds in corrugated plasma sheets, and ESEs are produced from \(A_3\) cusp catastrophes, which are stable singularities created as the ends of folded sheets meet \citep{10.1093/mnras/stae300}. This double catastrophe model has been used to explain a double lensing ESE, in which scattering is produced from the ESE source and the main scattering screen \citep{Zhu_2023}. Antisymmetric plasma lenses have been used to model ESE light curves and are able to reproduce the diverse observed event morphologies \citep{Dong_2018}. Progress on ESE models and investigations with volumetric ray tracing are explored in \cite{Au_2024}.

In this paper, we report evidence of an ESE detected toward PSR J2313+4253 (B2310+42). In Section \ref{sec:Data}, we discuss data acquisition. In Section \ref{sec:Analyses}, we discuss data analyses. In Section \ref{sec:results}, we present evidence for the ESE. Here we will include changes seen in the measured scintillation parameters and the estimated scattering screen distance. In Section \ref{sec:conlcusion}, we discuss the conclusions of this work.

\section{Data} \label{sec:Data}

Our observations were taken with the Green Bank Observatory 20m telescope as part of the Pulsar Science Collaboratory (PSC) Scintillation Survey \citep{PSCTurner_2024}. The PSC was created in 2007 and provides research opportunities that include pulsar searches, giant pulse studies, fast radio bursts, pulsar timing, and pulsar scintillation \citep{PSC,PSC2,Doskoch_2024}.  Each of our observations was 2 hours in length with the ``all" filter enabled, giving a frequency range of 1300--1800 MHz. This filter allows for the widest possible observing bandwidth, which provides tighter constraints on scintillation parameter estimates. For analysis, this range was subsequently cropped to 1350--1525 MHz, which is the maximum continuous frequency range with no significant narrow-band RFI. The observations ranged from MJD 60006--60783, which spans 777 days. We observed with a near-biweekly cadence. This was increased to a near-weakly cadence when it became apparent that a potential ESE was occurring. The sampling time was 4.18 ms and ephemerides were taken from the Australia National Telescope Facility (ATNF) pulsar catalog \citep{ATNF2005AJ....129.1993M}. 

In order to remove radio frequency interference (RFI), excision was performed using a custom median-smoothed zapping script, as well as manual removal of individual pixels. Once RFI was removed, the data spanned around 170 MHz centered near 1435 MHz. After folding and summing the data over polarizations, depending on the signal-to-noise ratio (S/N) of each observation, our data was averaged in frequency to yield channels of either 512 and 1024 frequency channels (approximately channel sizes of 1 or 0.5 MHz, respectively) across the observing band. The data was averaged in time with subintegrations ranging from 10--30 seconds, with 512 channels being used for particularly low S/N observations, in order to complete subsequent analyses. In Table \ref{tab:psrpar}, we provide a summary of the parameters for J2313+4253 with pulsar parameters taken from \cite{ATNF2005AJ....129.1993M} and scintillation parameter values from \cite{PSCTurner_2024}.

\begin{deluxetable*}{ccccccc}
\tablecaption{Pulsar Parameters \label{tab:psrpar}}
\tablehead{
\colhead{Pulsar} & \colhead{P} & \colhead{DM} & \colhead{$D_p$} & \colhead{Binary} & \colhead{$\overline{\Delta{\nu}_{d;1400}}$} & \colhead{$\overline{\Delta{t}_{d;1400}}$} \\
\colhead{J2000 Epoch} & \colhead{(seconds)} & \colhead{($\rm{pc \ cm^{-3}}$)} & \colhead{(kpc)} & \colhead{type} & \colhead{(MHz)} & \colhead{(minutes)}
}
\startdata
J2313+4253 & 0.349 & 17.28 & 1.060 & isolated & $5.2\pm1.6$ & $16.2\pm 2.6$ \\
\enddata
\label{tab:psrpar}
\end{deluxetable*}

\section{Analyses} \label{sec:Analyses}

Our scintillation analyses of this pulsar are the same as used in the main PSC scintillation study \citep{PSCTurner_2024}. The scintillation analysis for each observation was performed using \textsc{pypulse} \citep{pypulse} to extract a dynamic spectrum, given by the equation,

\begin{equation}
    S({\nu},{t})= \frac{P_{\rm{on}}(\nu,t)-P_{\rm{off}}(\nu,t)}{P_{\rm{bandpass}}(\nu,t)}
\label{eq:2}
\end{equation}

\noindent where $S$ is the intensity of the pulsar signal at each observing frequency, $\nu$, and time, t. $P_{\rm{bandpass}}$ is the total power of the observation, and $P_{\rm{on}}$ and $P_{\rm{off}}$  indicate the power in the on or off pulse components at each $\nu$ and t. To define an on-pulse window, a reference pulse was created using the template-generating algorithm in \textsc{pypulse}’s SinglePulse set of functions. Next, a secondary spectrum for each observation was generated by
 taking the squared modulus of the two-dimensional Fourier
 transform of the corresponding dynamic spectrum.

\subsection{Scintillation Bandwidth and Timescale} \label{subsec:scintband}

In order to determine the scintillation bandwidth, $\Delta \nu_{\rm{d}}$, and scintillation timescale, $\Delta t_{\rm{d}}$, we extracted a 2D autocorrelation function (ACF) using \textsc{pypulse} \citep{pypulse}, from each epoch's dynamic spectrum. With a summation of the 2D ACF over time and frequency, we created 1D ACFs in frequency and time, respectively. In this study, both of these 1D ACFs were fit with a Lorentzian to estimate the half width at half maximum (HWHM) for the frequency ACF to estimate scintillation bandwidth, and the half width at $1/e$ for the time ACF to estimate scintillation timescale \citep{1985ApJ...288..221C}. To determine the half width at $1/e$ of our time ACF, we multiplied the HWHM value by $\sqrt{e-1}$.

We are using a Lorentzian fit for our ACF, as opposed to many studies that use Gaussian \citep{Liu2022,wu2022}. This choice follows from the main PSC scintillation study \citep{PSCTurner_2024}, which demonstrated that the Lorentzian function was a better fit to dynamic spectrum ACFs. The ISM's pulse broadening function is a time domain function that exhibits the behavior of a one-sided decaying exponential. The frequency domain ACF is described by the Fourier pair of this function, which is a Lorentzian. To maintain consistency, we chose to fit the timescale with Lorentzian as well, which is often consistent with measurements using Gaussian fits within the uncertainty. 

\begin{figure}[htbp!]
\centering

\textbf{\includegraphics[scale=0.5] 
{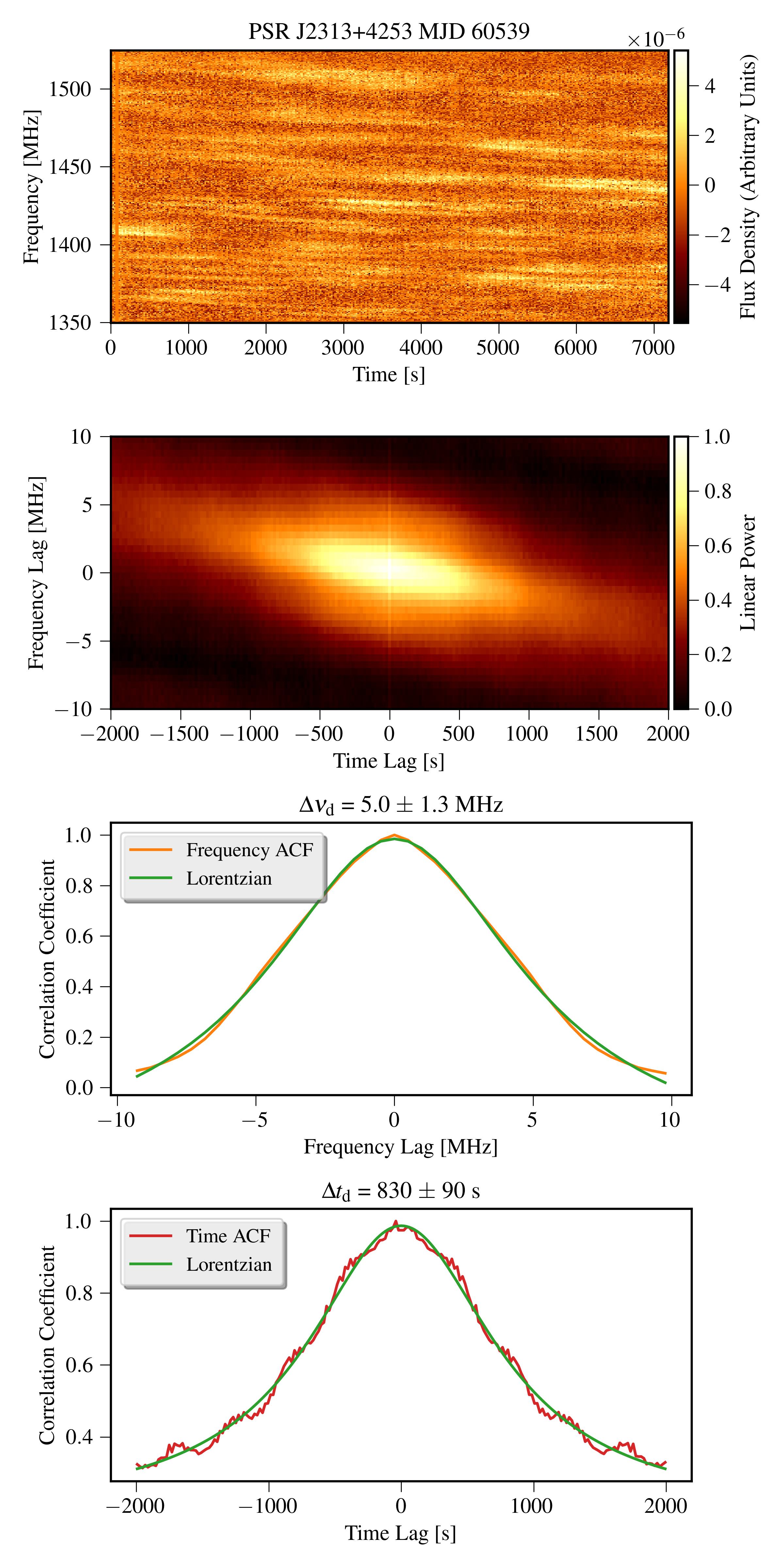}}

\caption{ A sample dynamic spectrum from MJD 60539 (top), the corresponding 2D ACF (second from top), and the 1D ACFs and fits in frequency (second from bottom) and time (bottom).
\label{fig:dynspec}}
\end{figure}

After excising RFI, the frequency range and central frequency of a given observation can vary compared to other epochs. To account for this variation, every measurement was scaled to its equivalent value at 1400 MHz before any analysis that used multiple measurements of the scintillation parameters was completed. This was carried out assuming that these lines of sight evolved in observing frequency following a Kolmogorov wave number spectrum, i.e., $\propto\nu^{4.4}$ for scintillation bandwidth and $\propto\nu^{1.2}$ for scintillation timescale. If there is any deviation from a Kolmogorov spectrum along this particular line of sight, we believe that any additional error introduced should be relatively small due to our small frequency range. A sample dynamic spectrum, along with the corresponding 2D ACF and 1D ACF fits, is shown in Figure \ref{fig:dynspec}.

The main limitation on the precision of our scintillation parameter measurements is the number of scintles visible in the dynamic spectrum for each epoch. The resulting uncertainty on the scintillation bandwidth is thus dominated by the finite scintle effect, described by

\begin{equation}
\begin{gathered}
    \epsilon =\frac{\Delta\nu_{\rm{d}}}{2\ln(2)N_{\rm{scint}}^{1/2}} \\ \approx \frac{\Delta\nu_{\rm{d}}}{2\ln(2)[(1+\eta_{\rm{t}}T/\Delta t_{\rm{d}})(1+\eta_{\rm{\nu}}B/\Delta\nu_{\rm{d}}]^{1/2}} ,
\end{gathered}
\end{equation}

\noindent where $N_{\rm{scint}}$ is the number of scintles seen in the corresponding dynamic spectrum, \textit{T} is the total integration time, \textit{B} is the total bandwidth, and $\eta_{\rm{t}}$ and $\eta_{\rm{\nu}}$ are filling factors that range from 0.1 to 0.3. Here we have chosen to set both to 0.2, following from \cite{1986ApJ...311..183C,PSCTurner_2024,fillingfactor22}. The finite scintle error for scintillation timescale uses the same equation with the exchange of $\Delta\nu_{\rm{d}}$ with $\Delta t_{\rm{d}}$ and the removal of the ln(2) factor.

\nobreak The measurement of scintillation bandwidth and timescale provides a means through which we can estimate the distance to the dominant scattering screen in a given observation. Assuming that the transverse velocity of this pulsar, determined by proper motion, $V_{\rm{pm}}$, is equal to the velocity of the ISM along the LOS to the pulsar \citep{1995ApJ...451..717G,Nicastro}, $V_{\rm{ISS}}$, we can find an expression for the distance to the dominant scattering screen. The screen distance is found by rearranging a combination of the expressions in \cite{gupta94} and \cite{Cordes_1998} for a given transverse velocity,

\begin{equation}
     V_{\rm{ISS}}=A_{\rm{ISS}}\frac{\sqrt{\Delta\nu_{\rm{d,MHz}}D_{\rm{p,kpc}}x}}{\nu_{\rm{GHz}}\Delta t_{\rm{d,s}}},
\label{eq:4}
\end{equation}

\noindent where $D_{\rm{p,kpc}}$ is the pulsar's distance in kpc, $A_{\rm{ISS}}$ is a scale factor which varies depending on assumptions related to the geometry and uniformity of the medium, and $x={D_{\rm{s}}/(}{D_{\rm{p}}-D_{\rm{s}}})$, where $D_{\rm{s}}$ is the distance from Earth to the screen. A value of $x=1$ corresponds to a scattering screen halfway between the observer and the pulsar. Using the assumption of a thin screen and a Kolmogorov medium, we choose $A_{\rm{ISS}}= 2.53 \times 10^4$ km $\rm{s^{-1}}$ following from \cite{Cordes_1998}. We are using the values of $D_{\rm{p}} = 1.06 \pm 0.08 $ kpc and $V_{\rm{ISS}} = 125 \pm 10$ km/s for J2313+4253 measured using VLBI from \cite{Chatterjee_2009}. Since this pulsar has a high transverse velocity, the effects from the ISM velocity should introduce a negligible contribution \citep{gupta94,Nicastro}. We also assume the screen is isotropic and that Earth's orbital velocity is comparatively small With these assumptions, the expression can subsequently be used to estimate scattering screen distance, $D_{\rm{s}}$.

\subsection{Scintillation Arcs} \label{subsec:Arcs}

\begin{figure}[ht!]
\centering

\textbf{\includegraphics[scale=0.4] 
{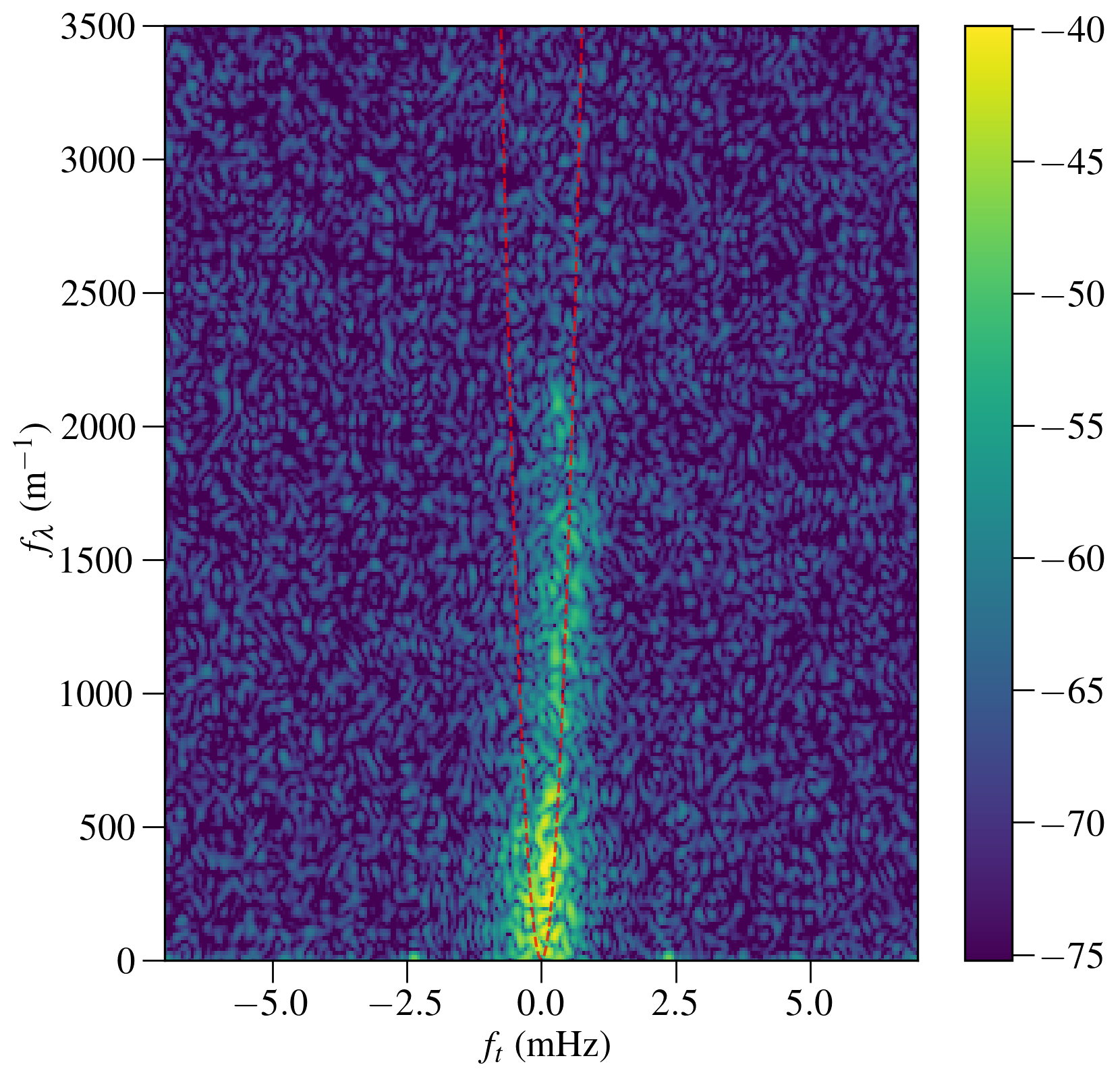}}

\caption{A scintillation arc and its corresponding fit (red) from an observation on MJD 60539. This epoch is after the ESE's conclusion. The color bar represents the logarithmic power in units of dB.}
\label{fig:arcfit}
\end{figure}

In addition to the approach described above, we also estimated scattering screen distances using scintillation arcs in the secondary spectra. For each epoch where an arc was resolved, arc curvature $\eta$ was calculated using the Hough transform technique provided in the \textsc{scintools} package \citep{Reardon_2020}. An example arc and its curvature fit are shown in Figure \ref{fig:arcfit}.

The curvature value of the arc is given by 

\begin{equation}
     \eta=\frac{\lambda^2}{2c}\frac{D_{\rm{eff}}}{V_{\rm{eff}}^2} ,
\label{eq:5}
\end{equation}

\noindent where $\lambda$ is the wavelength of the observation and $D_{\rm{eff}}$ is the effective distance, defined as ${D_{\rm{p}}D_{\rm{s}}}/({D_{\rm{p}}-D_{\rm{s}})}$ \citep{McKee_2022}. $V_{\rm{eff}}$ is the pulsar's effective velocity, and is given by

\begin{equation}
     V_{\rm{eff}} = \frac{1}{s}V_{\rm{ISS}}-\frac{1-s}{s}V_{\rm{pulsar}}-V_{\rm{Earth}},
\label{eq:6}
\end{equation}

\noindent where $s=1-{D_{\rm{s}}}/{D_{\rm{p}}}$ \citep{Stinebring_2001,Cordes_2006,McKee_2022}

Assuming that the velocity of the pulsar, which is over 100 km/s, is the dominant velocity in the system, then $V_{\rm{pulsar}}=V_{\rm{ISS}}$ and it follows that $V_{\rm{eff}} = V_{\rm{pulsar}}$. Also, because we do not know the screen orientation for this pulsar, we use the simplification that the angle between the major axis of the image on the sky and the system's effective velocity $\theta=0^{\circ}$. With this assumption, the calculated screen distance values should be treated as lower limits.

\section{Results} \label{sec:results}

\subsection{Scintillation Parameters} \label{subsec:Sband}

\begin{figure}[ht!]
\centering
{\includegraphics[width=7cm,height=5cm]{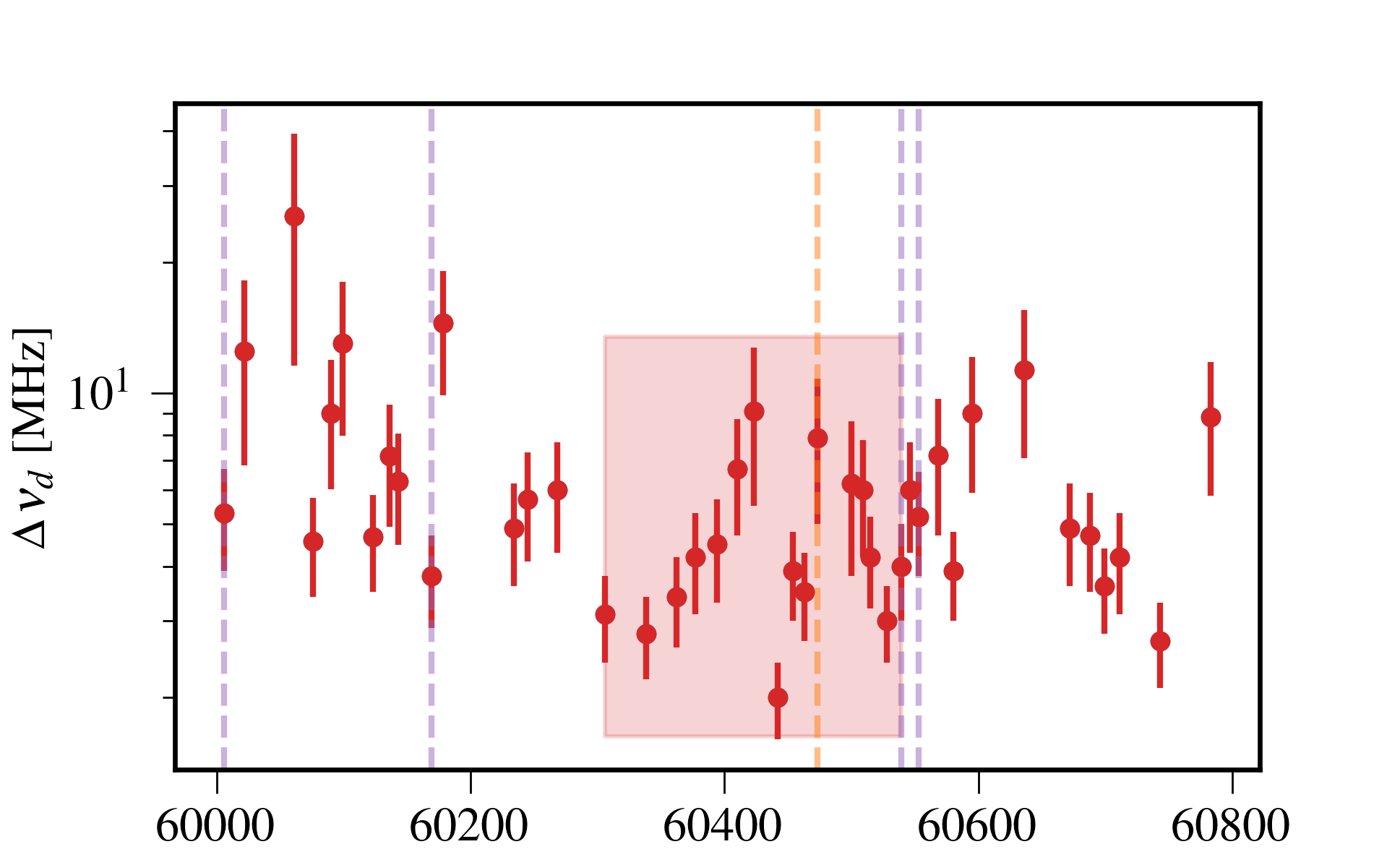}}
{\includegraphics[width=7cm,height=5cm]{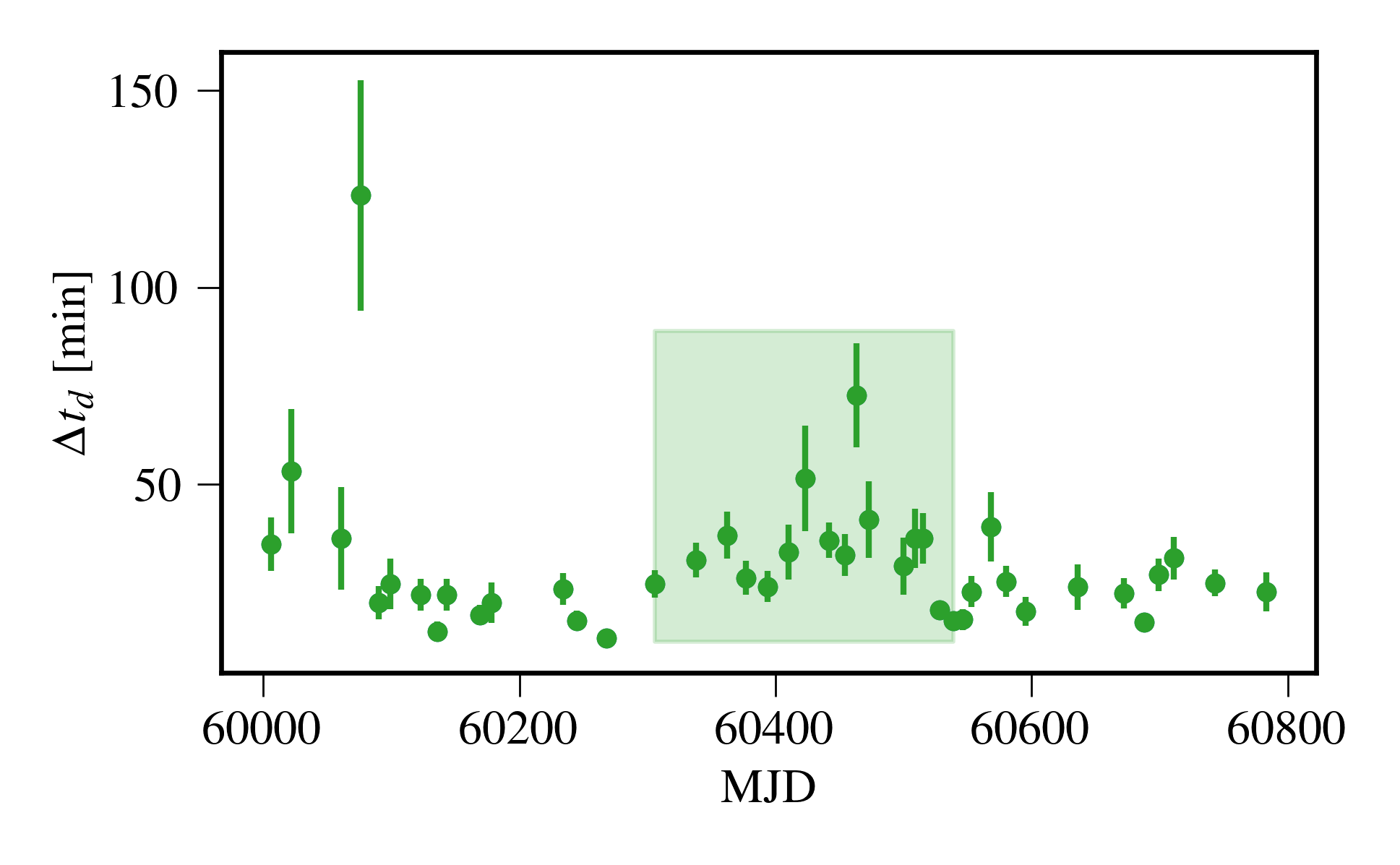}}
\caption{A time-series of the measured scintillation bandwidths $\Delta \nu_{\rm{d}}$ (top) and timescales $\Delta t_{\rm{d}}$ (bottom). The shaded regions indicate the duration over which the ESE takes place. The scintillation bandwidth values marked with vertical lines show the epochs for which a resolved arc was seen. The purple dashed lines are from epochs not during the event. The orange dashed line is during the event where the detached feature is seen.  
\label{fig:1}}
\end{figure}

In Figure \ref{fig:1}, we present a time series of the calculated scintillation bandwidths and scintillation timescales for each of our observations. The error bars are set at $1\sigma$. Also marked are the epochs for which scintillation arcs could be resolved. Before the event, the value of $\Delta \nu_{\rm{d}}$ fluctuates randomly around a mean value. The ESE is first seen around MJD 60306, where there is a sharp drop in $\Delta \nu_{\rm{d}}$. The scintillation bandwidth then rises to a spike, followed by another drop. The event then produces another spike and continues to fall until MJD 60528. Additionally, there is a simultaneous increase in $\Delta t_{\rm{d}}$ during the same period. We analyzed this trend using a sliding window t-test, which uses a fixed-size window that moves across the dataset, and computes independent t-statistics for each interval. Several window sizes from 12--20 data points were tested, with a window size of 16 points providing a balance between statistical stability and sensitivity to localized structure. After this test, with a fixed window size of 16, the boxed window was identified as the location showing the strongest statistical contrast versus unboxed region; this yields a p-value of 0.18 for scintillation timescale and 0.012 for scintillation bandwidth. A p-value less than 0.05 is considered significant, meaning the scintillation bandwidth p-value is  statistically significant, and the values are drawn from different distributions. The p-value for scintillation timescale is not significant. 

Using a t-test assumes that the values for each sample are independently drawn from a stationary distribution. Because our observations form a time series, any autocorrelation between samples could reduce the effective number of independent samples and increase significance levels \citep{hogg2010,chat}. To test for this effect, an autocorrelation function (ACF) was used and found a weak autocorrelation suggesting the values are only weakly dependent on previous measurements. The effective number of samples, using lags, is given by,

\begin{equation}
    n_{\rm eff}=\frac{N}{1+2\sum_{t=1}^{N-1}\rho_{t}},
\end{equation}

\noindent where $\rho_{k}$ is the autocorrelation at lag $k$ \citep{priestley1981spectral}. We found $n_{\rm eff}\approx31$ for the scintillation bandwidth measurements using lags up until the first sign change for $\rho_{t}$. This value remains close to the actual sample size $n=42$, suggesting that the dependence modestly reduces the degrees of freedom but does not qualitatively alter the inferences drawn from the test.

The 20m telescope currently does not allow for flux calibration during observations. As a result, we do not have flux density measurements; however, flux density is related to scintillation bandwidth. For diffractive scintillation, a decrease in the scattering strength correlates with an increase in the flux density as more light is directed toward the observer. In addition, scattering delay is inversely proportional to scintillation bandwidth. This means that a decrease in scattering strength correlates with an increase in scintillation bandwidth. As a result, we expect the timeseries of scintillation bandwidth and flux destiny to exhibit similar behavior. As a consequence of our high-cadence observations, we can observe the evolution of the ESE in scintillation bandwidth with more detail than in previous works. 

\subsection{Refractive Timescale} \label{subsec:ssepc}

Extreme scattering events are likely induced by refractive lensing. As a result, an ESE may correspond with a
sudden change in refractive timescale measured along the LOS. The refractive timescale can be calculated using the scintillation bandwidth and timescale, and is approximated by \citep{1990ApJ_refractive}, 

\begin{equation}
    t_r\approx\frac{4}{\pi}\Big(\frac{\nu\Delta\nu_{d}}{\Delta t_{d}}\Big).
\end{equation}

\begin{figure}
    \centering
    \includegraphics[width=7cm,height=5cm]{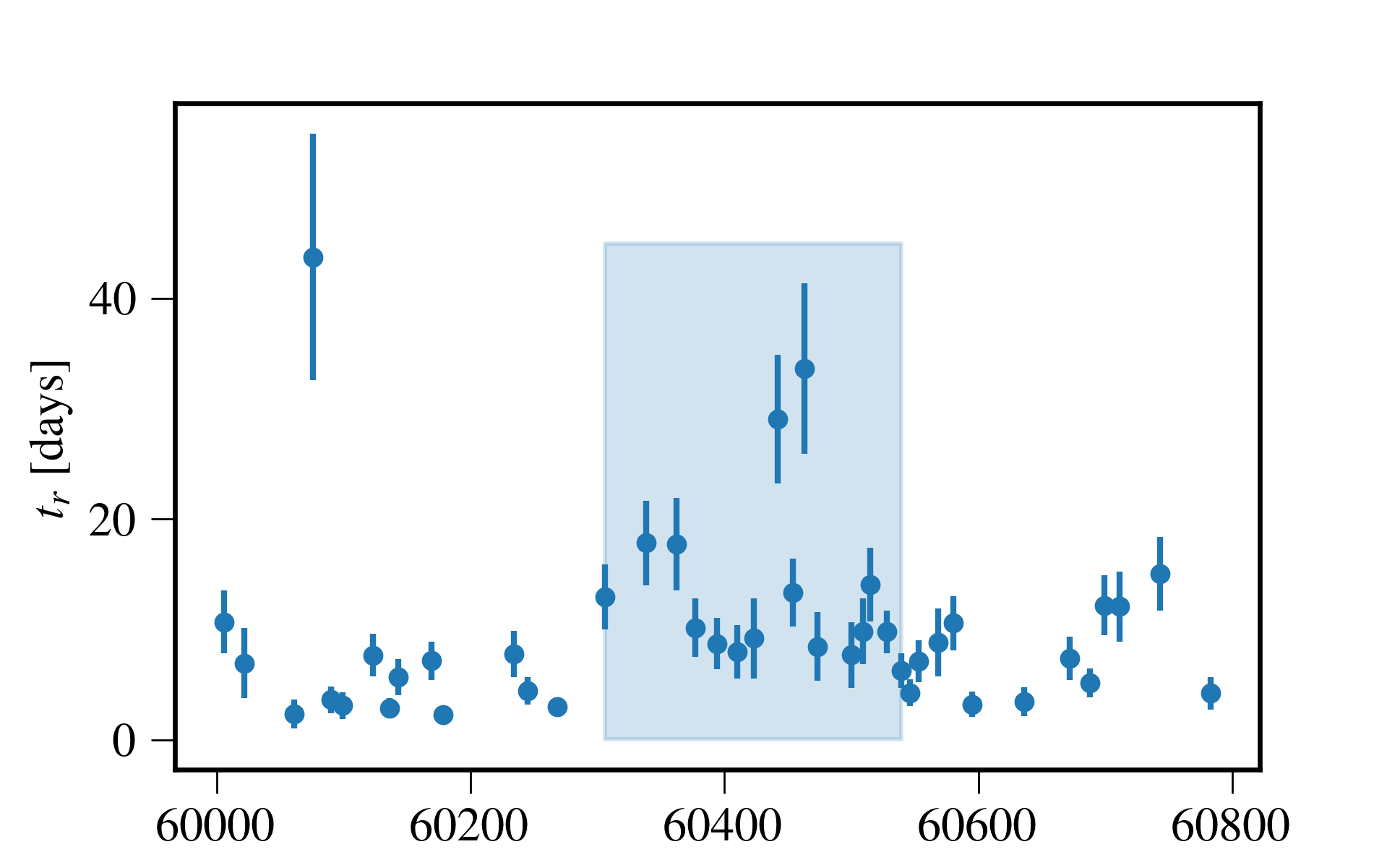}
    \caption{The calculated refractive timescale for each observation. The boxed region is the ESE event.}
    \label{fig:refractive}
\end{figure}

\noindent The results of this formula for each observation are shown in Figure \ref{fig:refractive}. There are two outliers in refractive timescale that correspond to the dip in scintillation bandwidth observed during the event. These occur on MJDs 60442 and 60463. Using the same t-test used in Section \ref{subsec:scintband}, for the refractive timescales in the same window, we find a p-value of 0.02, which is
statistically significant.

\subsection{ESE Source Distance and Scale} 
\label{subsec:esesource}

\begin{figure}[ht]
\centering
\includegraphics[width=7cm,height=5cm] {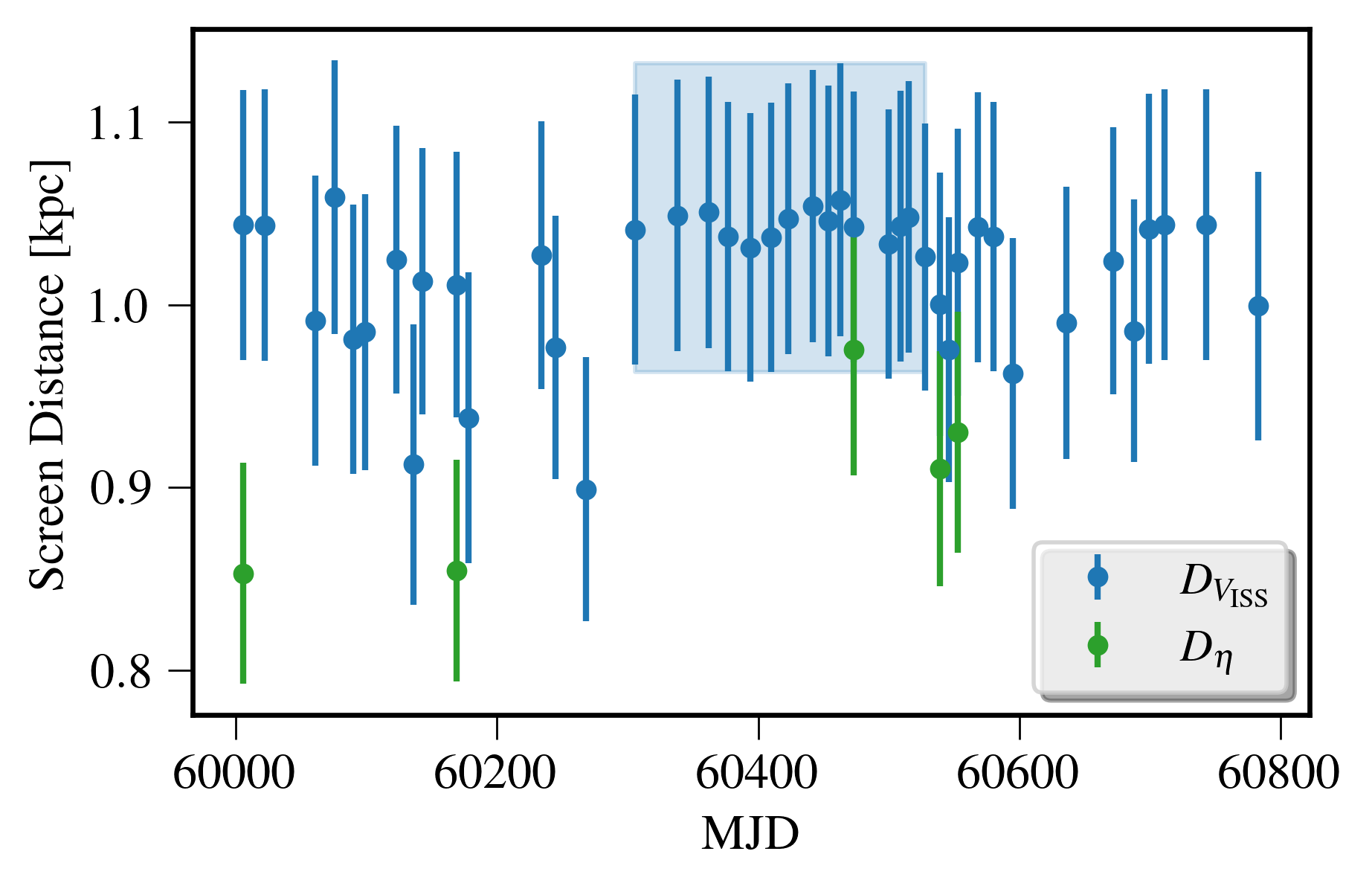}
\caption{A time-series of the calculated scattering screen distance. $ D_{V_{\rm ISS}}$ is calculated using the Lorentzian values for scintillation bandwidth and timescale (top). $ D_{\rm \eta}$ is calculated using the scintillation arc curvature. The ESE takes place in the region shaded in blue.
\label{fig:2}}
\end{figure}

In Figure \ref{fig:2}, we present a time-series of the calculated scattering screen distance using scintillation bandwidths and timescales $D_{V_{\rm{ISS}}}$ (blue) and arc curvature $ D_{\rm{\eta}}$ (green). There is a jump from MJD 60306--60528 in the calculated screen distance and the value is more tightly constrained, coinciding with the pattern observed in the scintillation bandwidth. The large error bars for the screen distance are dominated by the large pulsar distance uncertainty. The derived scattering screen distances vary on scales much smaller than their errors, indicating that their errors are overestimated. The weighted standard deviation of $D_{V_{\rm ISS}}$ values in the box is 0.014 kpc versus 0.042 kpc for the values outside the boxed region. This increase in screen distance is seen in the screen distance calculated using the dynamic spectrum scintillation parameters, $ D_{V_{\rm ISS}}$. The screen distance calculated using the scintillation arc curvature, $D_{\eta}$, is determined by fitting the main scintillation arc, which excludes the detached feature that appears during the ESE. This screen distance is consistent with measurements before and after the event. The $ D_{V_{\rm ISS}}$ screen distance before and after the event is less constrained, with a weighted average screen distance $ D_{V_{\rm ISS}} =1.00\pm0.04$ kpc, compared to a measured distance of $ D_{V_{\rm ISS}} = 1.04\pm0.01 $ kpc during the event. Using the same t-test used in Section \ref{subsec:Sband} on the $ D_{V_{\rm ISS}}$ values in the boxed region versus the unboxed region yields a p-value of 0.0006, which is very statistically significant.

Based on our scintillation bandwidth measurements, we estimate this ESE lasted approximately 220 days. Assuming the pulsar's velocity is the dominant in the system, then the pulsar will have moved by a distance equal to the duration of the event multiplied by the transverse velocity. Using some trigonometry and the small angle approximation, we can find an estimate of the transverse size of the structure the ESE originates from, scaled using the ratio of the screen distance and pulsar distance. While a more accurate estimate of the size of the structure that caused the ESE would require a more in-depth examination of the multipath nature of lens-pulsar emission interactions, we can make an approximation of the lens size, to first order, via the expression

\begin{equation}
    {L_{\rm{ESE}}={\frac{D_{\rm{s}}V_{\rm{eff}}t}{D_{\rm{p}}}}},
\label{eq:size}
\end{equation}

\noindent where $t$ is the duration of the event, ${V_{\rm{eff}}}$ is the effective velocity of the pulsar, and $D_{\rm{s}}$ and $D_{\rm{p}}$ are the distance to the screen and pulsar, respectively. This expression yields a value of $L_{\rm{ESE}} = $15.6$\pm 1.6$ AU. This agrees with \cite{2015ApJ...808..113C}, where the estimated size was on the order of AU. The agreement of these size estimates suggests that these events likely have similar structural origins.

\subsection{Secondary Spectra} \label{subsec:ssepc}
\begin{figure}[htbp!]
\centering
{\includegraphics[width=8cm,height=5.2cm]{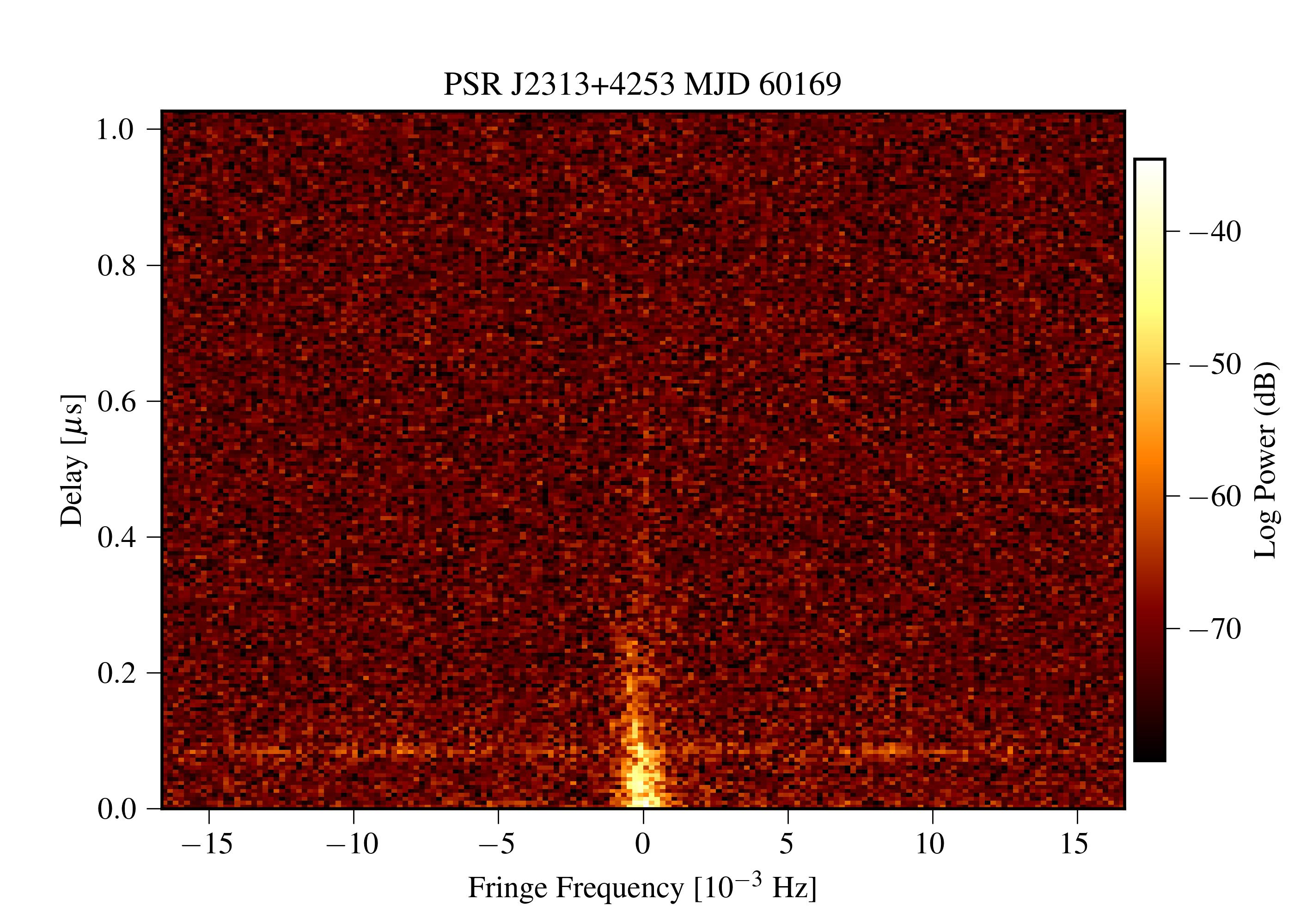}}
{\includegraphics[width=8cm,height=5.2cm]{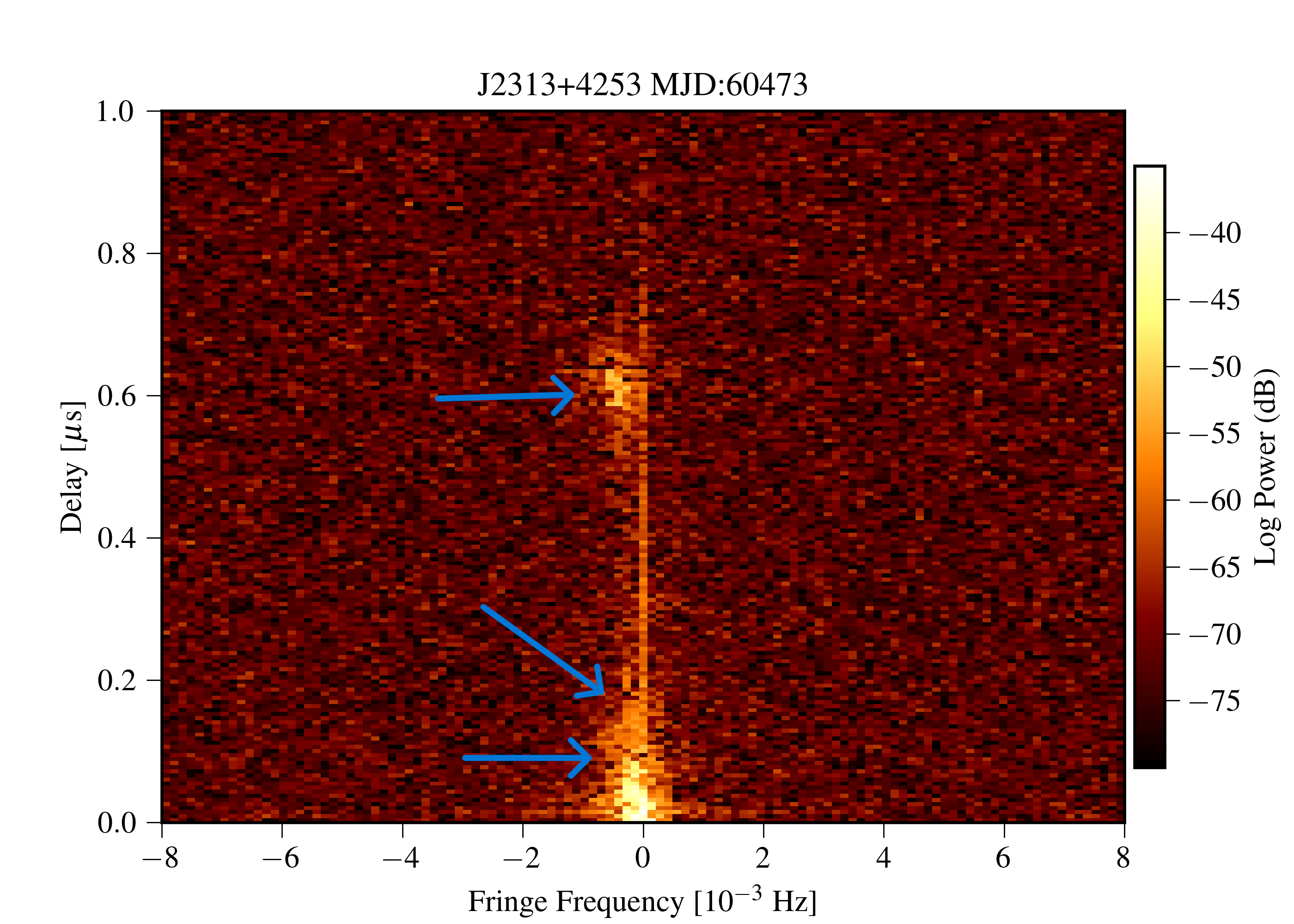}}
{\includegraphics[width=8cm,height=5.2cm]{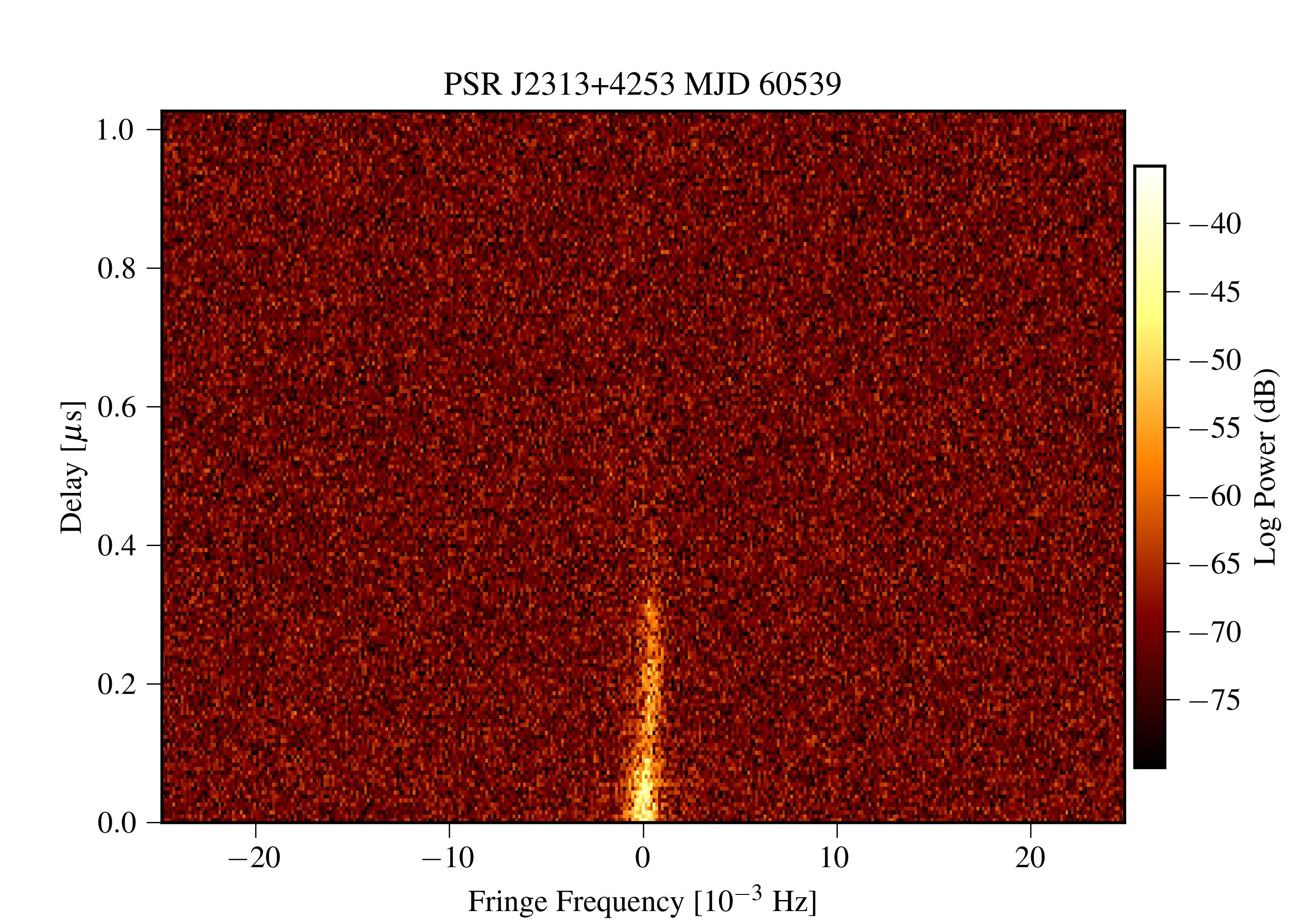}}
\caption{A secondary spectrum corresponding to an observation before (top), during (middle), and after (bottom) the ESE. In the observation taken during the event, there is a detached feature from the main arc corresponding to a delay value of 0.6$ \mu s$ which is not present in the secondary spectra processed before or after the event. This detached feature resembles those seen in other double lensing events. The horizontal line around 0.08 $ \mu s$ in the top panel is likely an artifact from RFI. The points used for the angular scale calculation are marked with blue arrows. The top, middle, and bottom arrows correspond to points A, B, and C, respectively.} 
\label{fig:secondayevent}
\end{figure}

\begin{figure}
    \centering
    \includegraphics[scale=0.4]{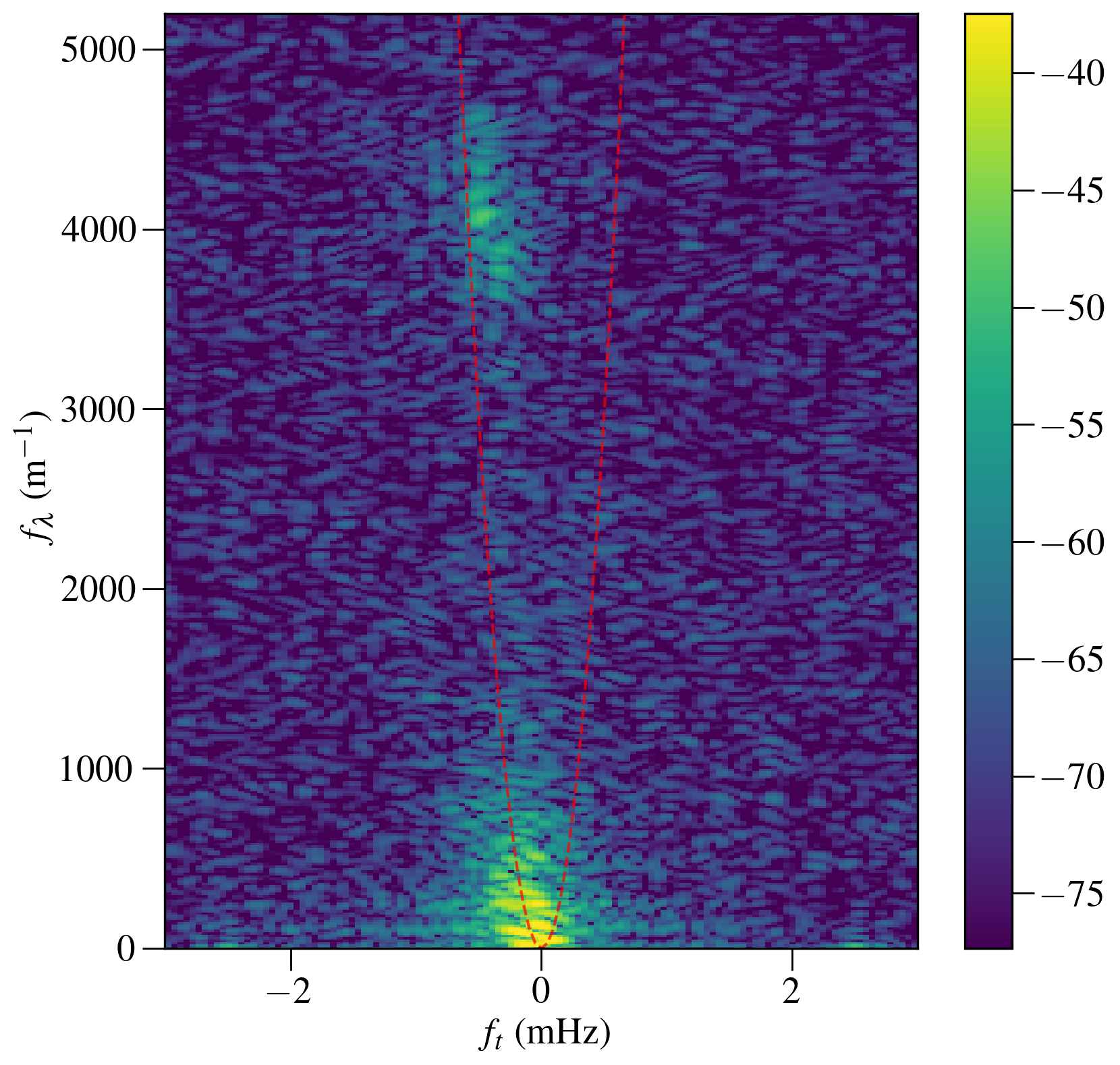}
    \caption{The secondary spectrum along with the corresponding curvature fit for MJD 60473.}
    \label{fig:esearcfit}
\end{figure}

In Figure \ref{fig:secondayevent}, we present a series of secondary spectra observed before, during, and after the ESE. The spectrum from an observation during the event contains a detached feature with a significantly larger delay value than the rest of the main scintillation arc. This feature was only seen during the event and does not appear in any secondary spectra from before or after the event. For the detached feature at $0.6$ $\mu s$, images are created from photons scattered at angles around 1.4--1.7 times greater than those in the main arc, which terminates at $0.2$ $\mu s$. The main scintillation arc is created from the usual dominant scattering screen, while the detached feature is created by an additional separate structure, which is likely the origin of the ESE. The best fit parabola for the scintillation arc curvature corresponding to the secondary spectrum on MJD 60473 is shown in Figure \ref{fig:esearcfit}. Based on the screen distance estimates before and during the event, the light emitted from the pulsar is likely first scattered by this ESE lens before being scattered by the primary screen. A detached feature like this one seen in our secondary spectrum is indicative of a double-lensing event \citep{Zhu_2023}.

In the secondary spectrum, features with power at larger delays correspond to photons scattered at a larger angle. At the time this detached feature is seen, the ESE is at the peak of a cusp (dashed orange line). This means that the scattering is at its weakest and the contributing photons are originating from a minium scattering angle. At other times during the event, like during the sharp drop in the scintillation bandwidth, the contributing photons are scattered at much higher angles. During these epochs, due to our limited range in delay, caused by our finite frequency resolution, we strongly suspect the detached feature is at a delay above the maximum we can resolve and therefore not visible in our secondary spectra. This is similar to the detached 1 ms feature identified by \cite{2010ApJ...708..232Bbrisken} and analyzed by \cite{Zhu_2023}, which starts at a higher value on the delay axis and travels down the arc as the event progresses.

We can estimate the angular scale of features in our secondary spectrum, which is useful to determine the relative position of the main scattering screen and the structure the ESE originates from. The angular scale $\theta_{2}$ is found by rearranging Equation \ref{eq:1} \citep{Hill_2003}, 

\begin{equation}
    f_{\nu}=\left [\frac{D_{p}(D_{p}-D_{s})}{2cD_{p}}\right](\theta_{2}^{2}-\theta_{1}^{2}),
\label{eq:1}
\end{equation}

\noindent where $f_{\nu}$ is the conjugate frequency, $D_{p}$ is the distance to the pulsar, $D_{s}$ is the distance from the pulsar to the scattering screen, $\theta_{1}$ and $\theta_{2}$ are two points on the image in a coordinate system centered on the pulsar, and \textit{c} is the speed of light. We assume that the scattering is one dimensional and set $\theta_{1}=0$. This fixes one of the two points to the origin.

\begin{figure}[ht!]
\centering
{\includegraphics[width=8cm,height=5.2cm]{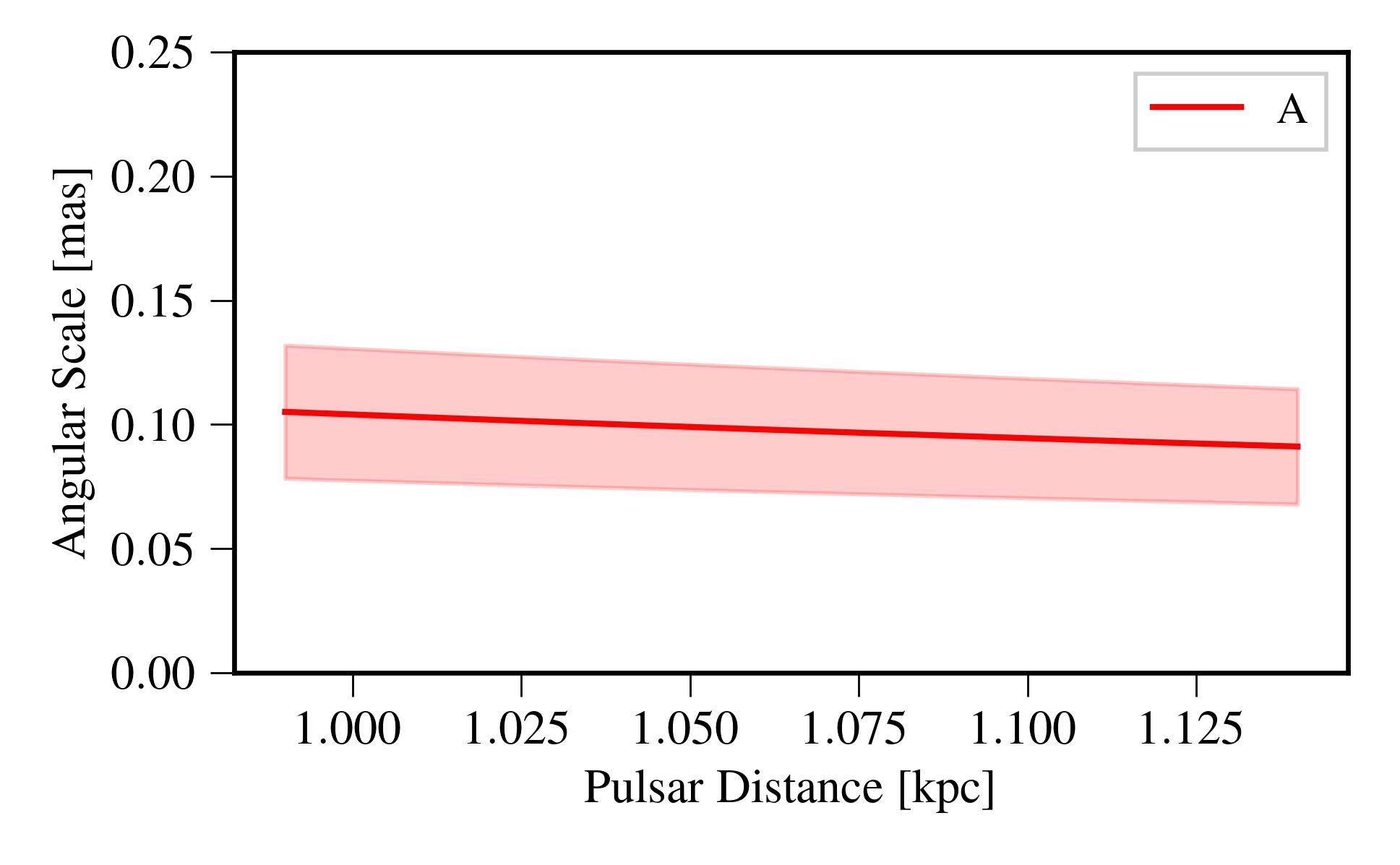}}
{\includegraphics[width=8cm,height=5.2cm]{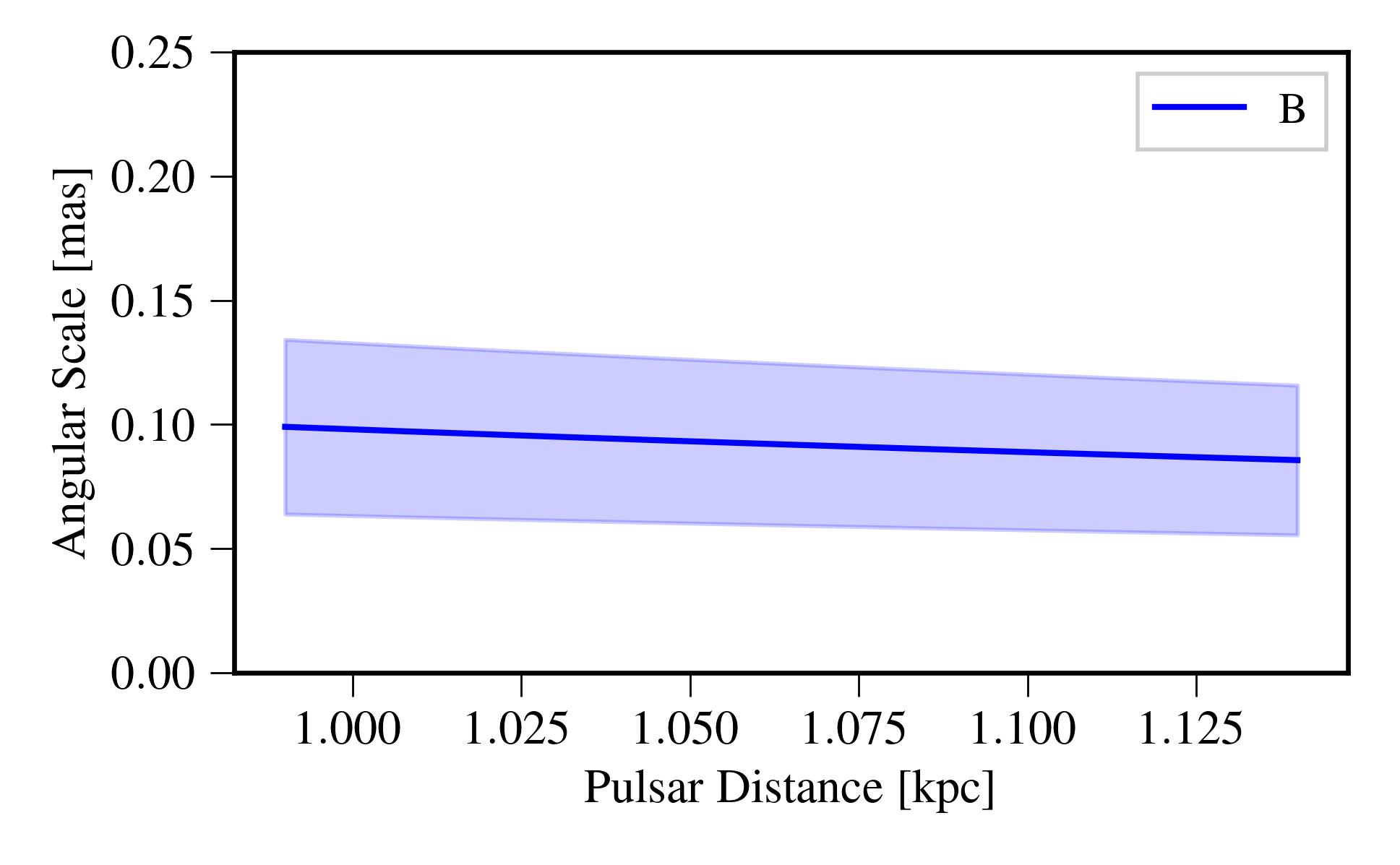}}
{\includegraphics[width=8cm,height=5.2cm]{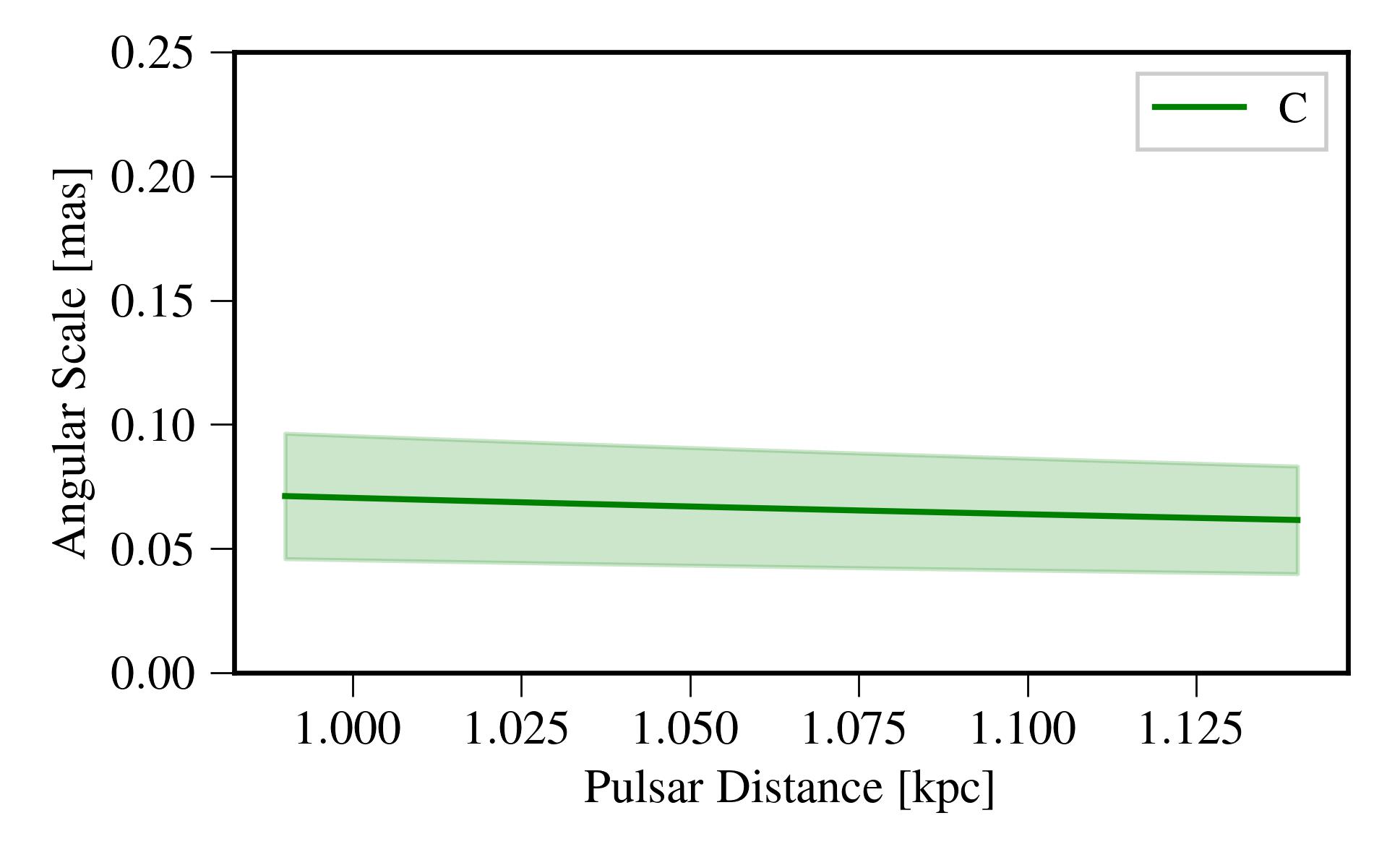}}
\caption{The calculated angular scale for features in the secondary spectrum corresponding to MJD 60473. (A) represents the detached feature. Points (B) and (C) are along the main scintillation arc. The calculated angular scale is shown over the range of all possible pulsar distances, based on the uncertainty on the estimated distance. The error for each is shown in the shaded region and is derived from the uncertainty on the pulsar distance. 
\label{fig:4}}
\end{figure}

In Figure \ref{fig:4}, we plot the results of Equation \ref{eq:1} for three points along the scintillation arc present in the middle plot shown in Figure \ref{fig:secondayevent} marked with blue arrows. Since there is uncertainty on the estimated pulsar distance, we calculate the angular scale for each possible distance based on the range given by the 1$\sigma$ error from \cite{Chatterjee_2009}. Point A represents the detached feature, with a maximum delay value of 0.6 $\mu s$. Points B and C are assigned to values along the main scintillation arc, with delay values of 0.2  $\mu s$ and 0.1  $\mu s$, respectively. Points B and C were chosen as the location of images that are on the main scintillation arc and are seen in every observation, with point B being at the maximum value of the main scintillation arc for MJD 60473. Point A is determined using the ESE screen distance from Section \ref{subsec:esesource}, $ D_{V_{\rm ISS}} = 1.04\pm0.01 $. Similarly, points B and C are found using the screen distance outside the event, $ D_{V_{\rm ISS}} =1.00\pm0.04$. The error for each angular scale is shown in the shaded region. The image at A corresponds to a higher angular scale value than the images at B and C.

\section{Conclusions \& Future Work} \label{sec:conlcusion}

We have presented evidence of an extreme scattering event in the direction of J2313+4253. We observed a characteristic drop in the scintillation bandwidth along with spikes, a trend that is similar to that seen in other ESEs. There is a jump in the estimated screen distance, which became more constrained during the event. In addition, we observe a detached
feature in one of our secondary spectra, which indicates scattering originating from a separate source than the main scattering screen, and is indicative of a double-lensing event. The ESE duration was approximately 220 days.  Assuming that the transverse velocity of this pulsar dominates the velocity of the diffraction pattern along the LOS, the estimated screen distance for the structural origin of the event is 1.04$\pm 0.01$ kpc and the transverse size is 15.6$\pm 1.6$ AU. This places the structure close to the pulsar, which is approximately $1.06$ kpc from Earth. The detached feature present in the secondary spectrum from an observation during the event had a maximum delay value of 0.6 $\mu s$, which corresponds to scattering angles 1.4--1.7 times greater than those in the main arc. 

High cadence observations of ESEs allow for measurement of extremely small scale features in the ISM, estimates of ISM velocities, and approximation of the densities of these features. This can improve our understanding of the small scale structures in the ISM. Additionally, sudden changes in the scattering along a LOS could create shape timing offsets, which introduce pulsar timing noise, impacting pulsar timing arrays.

With future monitoring of scintillation parameters for more pulsars, additional ESEs may be identified. ESEs toward pulsars can be identified by monitoring the scintillation parameter timeseries for statistical changes, looking for sudden changes in the refractive timescale, and analyzing secondary spectra for transient detached features. Telescopes that continuously monitor pulsars could be used as an ESE detection system. This would allow for a higher detection rate. Once an ESE is identified, triggered proposals could be used to gather observations at a higher cadence, allowing for more detailed analysis of these events.  This will help to further constrain and test theories for the origins of these events.

\begin{acknowledgments}
\textit{Acknowledgments}: This work was made possible by the Pulsar Science Collaboratory (PSC). Through the PSC we were granted regular access to the Green Bank 20m Telescope, without which this study would not have been possible. The Pulsar Science Collaboratory (PSC), a partnership between West Virginia University, the Green Bank Observatory, and the NANOGrav collaboration is funded by the National Science Foundation (awards: \#1516512, \#1516269, \#2020265). The National Radio Astronomy Observatory and Green Bank Observatory are facilities of the U.S. National Science Foundation operated under cooperative agreement by Associated Universities, Inc. Thanks to Victoria Catlett for valuable discussion regarding the visualization of Figure \ref{fig:4}.
\end{acknowledgments}

\vspace{5mm}

\software{\textsc{psrchive} \citep{psrchive}, \textsc{pypulse} \citep{pypulse}, \textsc{scintools} \citep{scintools}, \textsc{scipy} \citep{scipy}, \textsc{numpy} \citep{numpy}, \textsc{matplotlib} \citep{matplot}, \textsc{ultranest} \citep{buchner2021ultranestrobustgeneral}}


\bibliography{PSC_ESE}{}
\bibliographystyle{aasjournal}



\end{document}